\RequirePackage[2020-02-02]{latexrelease}
\documentclass[twocolumn]{revtex4}

\usepackage{graphicx}
\usepackage{amsmath}
\usepackage{amssymb}

\newcommand{\RT}[1]{``#1,''}
\newcommand{\AppExpSetup}{Sec.~I of the Supplemental Materials}
\newcommand{\AppTheory}{Sec.~II of the Supplemental Materials}
\newcommand{\AppSBR}{Sec.~III of the Supplemental Materials}
\newcommand{\AppEqs}{Eqs.~(S2)-(S4) of the Supplemental Materials}

\newcommand{\CiteSfwmDl}{
SFWM.DL.LCA.2005, 
SFWM.DL.LCA.2006, 
SFWM.DL.LCA.2008, 
SFWM.DL.LCA.2009, 
SFWM.DL.LCA.2010, 
SFWM.DL.LCA.2011, 
SFWM.DL.LCA.2014a, 
SFWM.DL.LCA.2014b, 
SFWM.DL.LCA.2015a, 
SFWM.DL.LCA.2015b, 
SFWM.DL.LCA.2016a, 
SFWM.DL.LCA.2016b, 
SFWM.DL.LCA.2018, 
SFWM.DL.LCA.2019a, 
SFWM.DL.LCA.2019b, 
SFWM.DL.LCA.2020, 
OurAPL2022, 
SFWM.DL.LCA.2023, 
YFCPRR2024, 
SFWM.DL.RHA.2016, 
SFWM.DL.RHA.2017a, 
SFWM.DL.RHA.2017b, 
SFWM.DL.RHA.2018, 
SFWM.DL.RHA.2019, 
SFWM.DL.RHA.2020a, 
SFWM.DL.RHA.2020b, 
SFWM.DL.RHA.2021, 
OurOPEX2021, 
SFWM.DL.RHA.2022, 
OurPRR2022, 
OurPRA2022, 
OurOPEX2024, 
OurQST2025}
\newcommand{\CiteBiTheory}{
BiphotonTheory.2008, 
OurAQT2024}
\newcommand{\CiteSpdc}{
SPDC.2012,
SPDC.2013,
SPDC.2015a,
SPDC.2015b,
SPDC.2016a,
SPDC.2016b,
SPDC.2017a, 
SPDC.2017b,
SPDC.2018,
SPDC.2019,
SPDC.2020}

\begin{document}

\title{Protecting Heralded Single Photons Generated from Double-$\Lambda$ Biphoton Sources with Doppler-Broadened Atomic Media}

\author{
Wei-Kai Huang,$^{1,}$\footnote{The two authors contributed equally.}
Tse-Yu Lin,$^{1,\ast}$
Pei-Yu Tu,$^{1}$
Yong-Fan Chen,$^{2,3}$
Ite A. Yu$^{1,3,}$}\email{yu@phys.nthu.edu.tw}

\affiliation{
$^{1}$Department of Physics, National Tsing Hua University, Hsinchu 30013, Taiwan \\
$^{2}$Department of Physics, National Cheng Kung University, Tainan 70101, Taiwan \\
$^{3}$Center for Quantum Science and Technology, National Tsing Hua University, Hsinchu 30013, Taiwan
}

\begin{abstract}
Biphoton sources that use room-temperature or hot atoms are valuable for real-world applications in long-distance quantum communication and photonic quantum computation. The heralded single photons produced by biphoton sources using the double-$\Lambda$ spontaneous four-wave mixing (SFWM) process offer advantages of narrow linewidth, stable frequency, and tunable linewidth---qualities not found in other types of biphoton sources. In this study, we investigated a hot-atom SFWM double-$\Lambda$ biphoton source. We discovered that, under the condition counterintuitive to the present theory, heralded single photons of the source enhanced their generation rate by a factor of 3.6, heralding probability by a factor of 3.0, temporal width by 2.1, and spectral brightness by a factor of 10. These unexpected findings led us to propose a new theoretical framework for a previously unexplored physical mechanism. Our proposed theory effectively explains the observed results. Traditionally, similar spectral brightness (SB) from atom-based sources resulted in a lower signal-to-background ratio (SBR) than crystal- or chip-based biphoton sources, mainly due to poorer heralding probabilities. In our work, we experimentally demonstrated that the SBR improved by a factor of 4.8 while maintaining a comparable SB. As a result, the SBR performance of the atom-based biphoton source is now on par with that of crystal- or chip-based sources. This research introduces a new tuning parameter for double-$\Lambda$ SFWM biphoton sources, enhances our understanding of biphoton generation, and opens new avenues for improving the performance of these sources.
\end{abstract}

\maketitle

\newcommand{\FigOne}{
\begin{figure}[t]
	\includegraphics[width=\columnwidth]{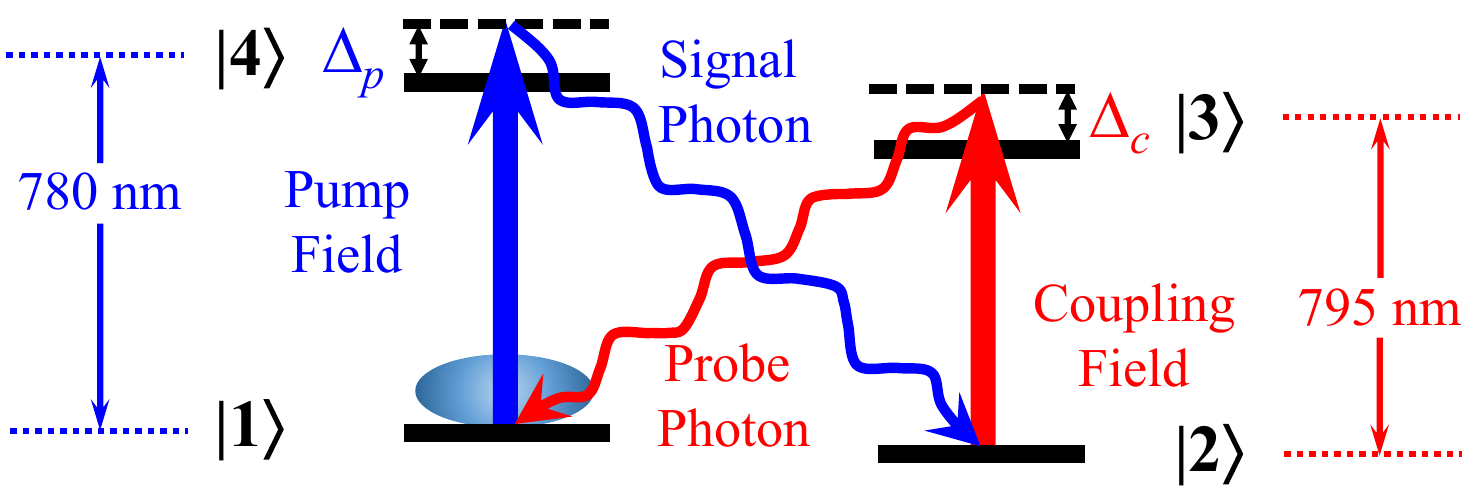}
	\caption{
Transition diagram of the double-$\Lambda$ spontaneous four-wave mixing process. We kept $\Delta_{p}$ to 1.9~GHz and varied $\Delta_c$ between 0 and 3.0~GHz.  In the experiment, the heated atomic vapor had the Doppler $e^{-1}$ half width of about 320~MHz for all the transitions from a ground state $|1\rangle$ or $|2\rangle$ to an excited state $|3\rangle$ or $|4\rangle$.
	}
	\label{fig:transition}
	\end{figure}
}
\newcommand{\FigTwo}{
	\begin{figure}[t]
	\includegraphics[width=\columnwidth]{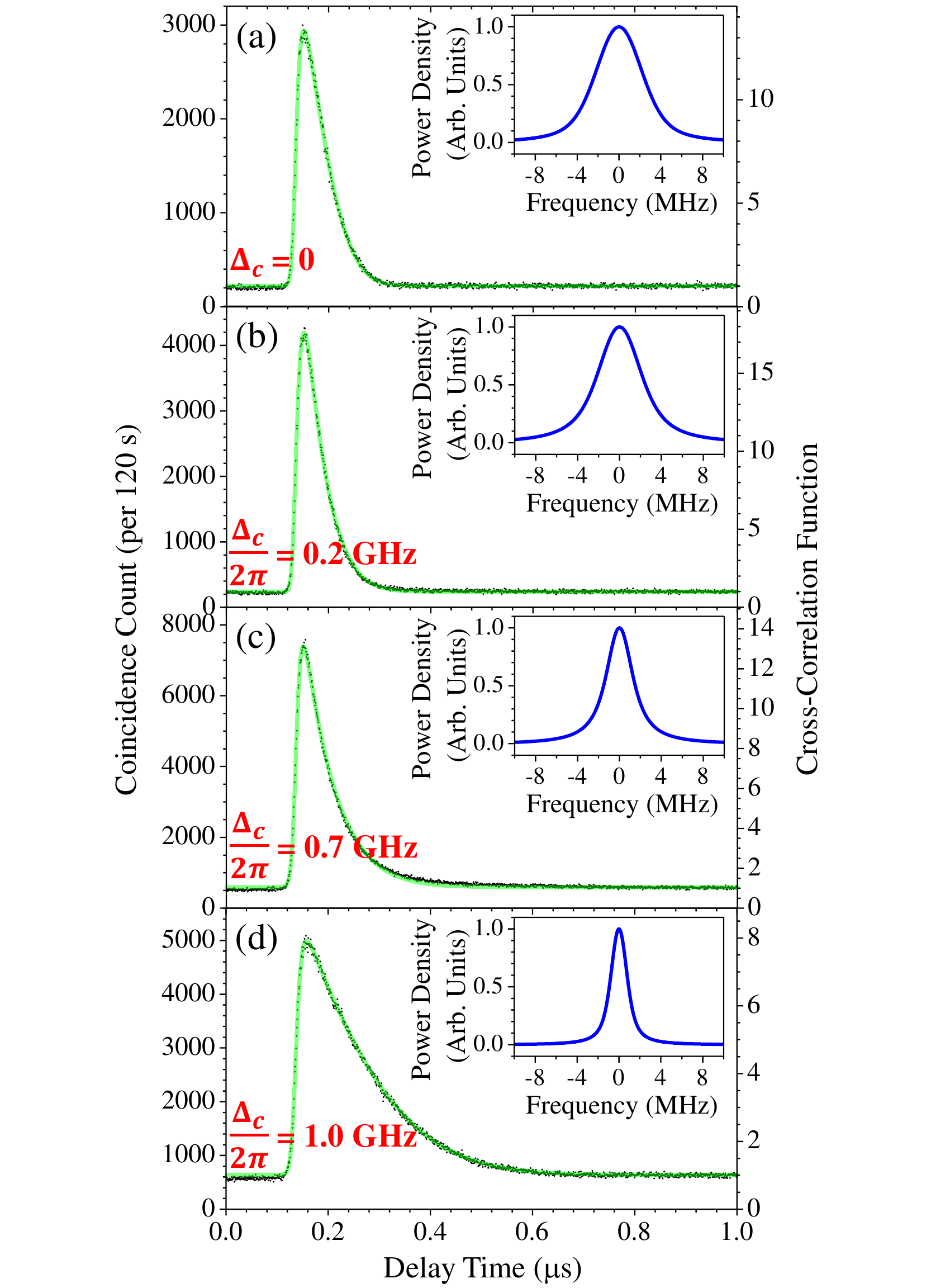}
	\caption{
Representative data showing coincidence count (left axis) and cross-correlation (right axis) versus the delay time between the signal and probe photons. Black dots represent the experimental data, and green lines are the best fits. Each inset shows the frequency spectrum of the best fit. We set the time bin width to 0.8~ns and the accumulation time to 120~s in the coincidence count measurement. During the measurement, the OD was 510; the pump detuning and power were 1.9~GHz and 5~mW; the coupling power was 17~mW. We list the coupling detuning and the results in Table~\ref{table:ForFigTwo}.
	}
	\label{fig:biphoton}
	\end{figure}
}
\newcommand{\TableForFigTwo}{
	\begin{table}[t]
	\caption{Results of the biphoton data in Fig.~\ref{fig:biphoton}. 
	}
{\centering
	\footnotesize
	\begin{tabular}{c c c c c c} 
	\hline 
	\parbox{12mm}{\vspace*{1mm}
		$\Delta_c/2\pi$\\ \vspace*{2pt}(GHz)
				\vspace*{1mm}} &
	\parbox{13mm}{$R_g$$^{\ast}$\\ \vspace*{2pt}($10^5$/s)} &
	\parbox{13mm}{$\tau_w$\\ \vspace*{2pt}(ns)} &
	\parbox{13mm}{$\Delta\omega/2\pi$\\ \vspace*{2pt}(MHz)} &
	\parbox{12mm}{SBR} &
	\parbox{14mm}{$h_p$$^{\ast}$} \\
 	\hline 
	0.0 & 1.79$\pm$0.03$^\dagger$ & 63.6$\pm$0.7 & 5.2$\pm$0.1 & 
		12.4$\pm$0.3 & ~26.2$\pm$0.4\%$^\ddagger$ \\ 
	0.2 & 2.32$\pm$0.01$^\dagger$ & 54.1$\pm$0.5 & 5.1$\pm$0.1 & 
		16.4$\pm$0.7 & 38.0$\pm$0.3\% \\
	0.7 & 5.35$\pm$0.09$^\dagger$ & 65$\pm$1 & 3.0$\pm$0.1 & 
		11.9$\pm$0.3 & 74$\pm$1\% \\ 
	1.0 & 6.42$\pm$0.03$^\dagger$ & 132$\pm$1 & 1.83$\pm$0.03 & 
		6.8$\pm$0.1 & 79.9$\pm$0.4\%$^\ddagger$ \\
	\hline
	\end{tabular}
	\hspace*{4.5pt}
	\parbox{87mm}{
		\vspace*{-2mm}
		\begin{flushleft}
		\hspace*{-5pt}$^{\ast}${Referring to the biphotons collected in the 
			polarization-maintained optical fibers.}\\
		\hspace*{-5pt}$^{\dagger}${Multiply each value by a factor of 1.9 to obtain 
			the $R_g$ right after the vapor cell.}\\
		\hspace*{-5pt}$^{\ddagger}$By using the probe photons as the heralding
			 photons \cite{OurQST2025}, this value becomes 
			58.8$\pm$0.3\% at 0~GHz or 86.4$\pm$0.3\% 
			at 1~GHz.\\
		\end{flushleft}
	}
}
	\label{table:ForFigTwo}
	\end{table}
}
\newcommand{\FigThree}{
	\begin{figure}[t]
	\includegraphics[width=\columnwidth]{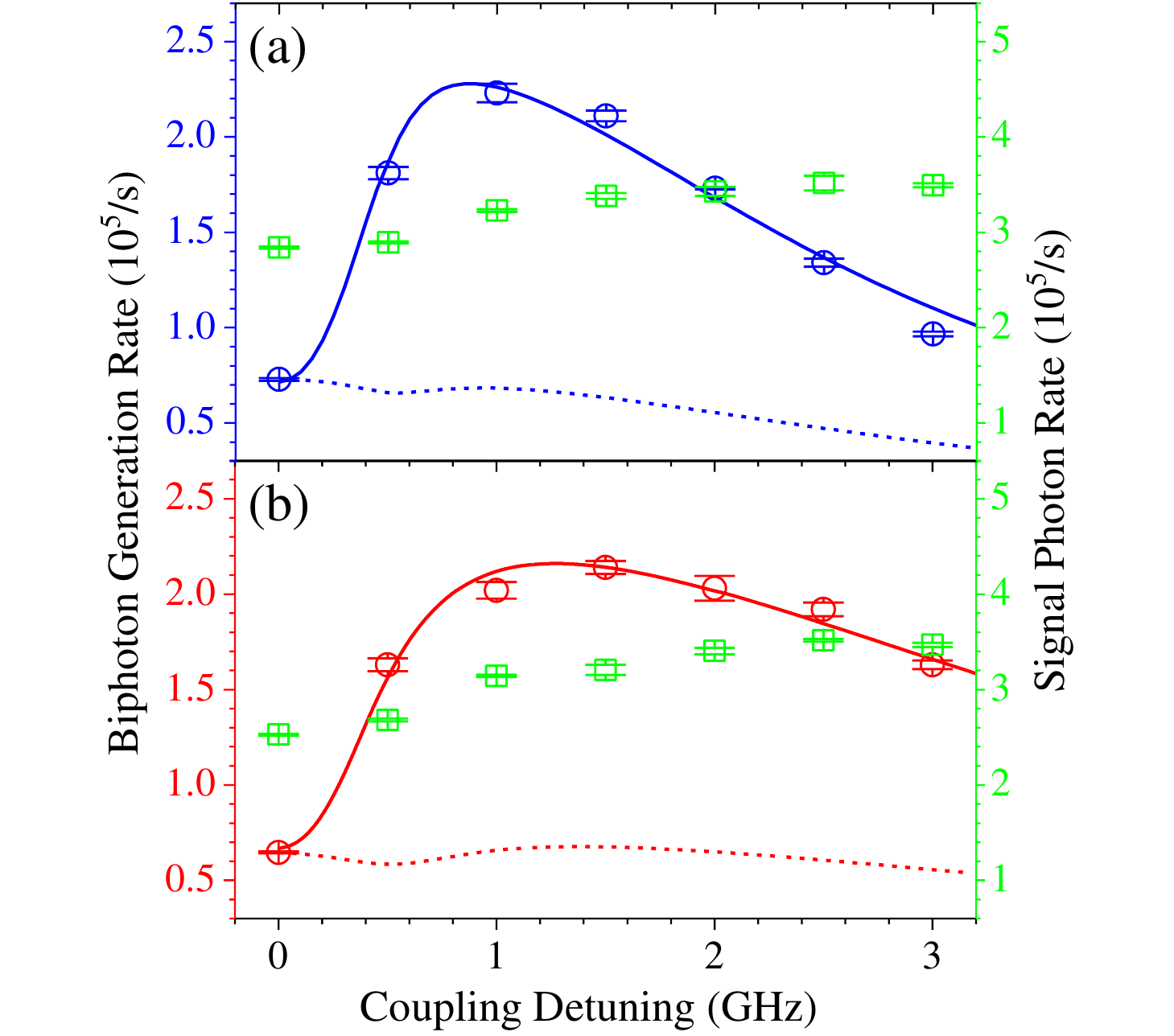}
	\caption{
The biphoton generation rate (left axis and blue or red circles) and signal-photon generation rate (right axis and green squares) are functions of the coupling detuning. Circles and squares represent the experimental data. In the experiment, the OD was 500$\pm$10, and the pump detuning and power were 1.9~GHz and 2~mW. Solid and dashed lines are the predictions based on the theoretical models with ($b \neq 0$) and without ($b = 0$) the impurity atoms, respectively. (a) We set the coupling power to 15~mW. The predictions were calculated with $\Omega_c$  $= 11.4\Gamma$, $\gamma$ $= 0.013\Gamma$, and $b$ $= 0.375$ in \AppEqs. (b) We set the coupling power to 30~mW. The predictions were calculated with $\Omega_c = 16.6\Gamma$, $\gamma = 0.010\Gamma$, and $b = 0.315$.
	}
	\label{fig:GR}
	\end{figure}
}
\newcommand{\FigFour}{
	\begin{figure}[t]
	\includegraphics[width=\columnwidth]{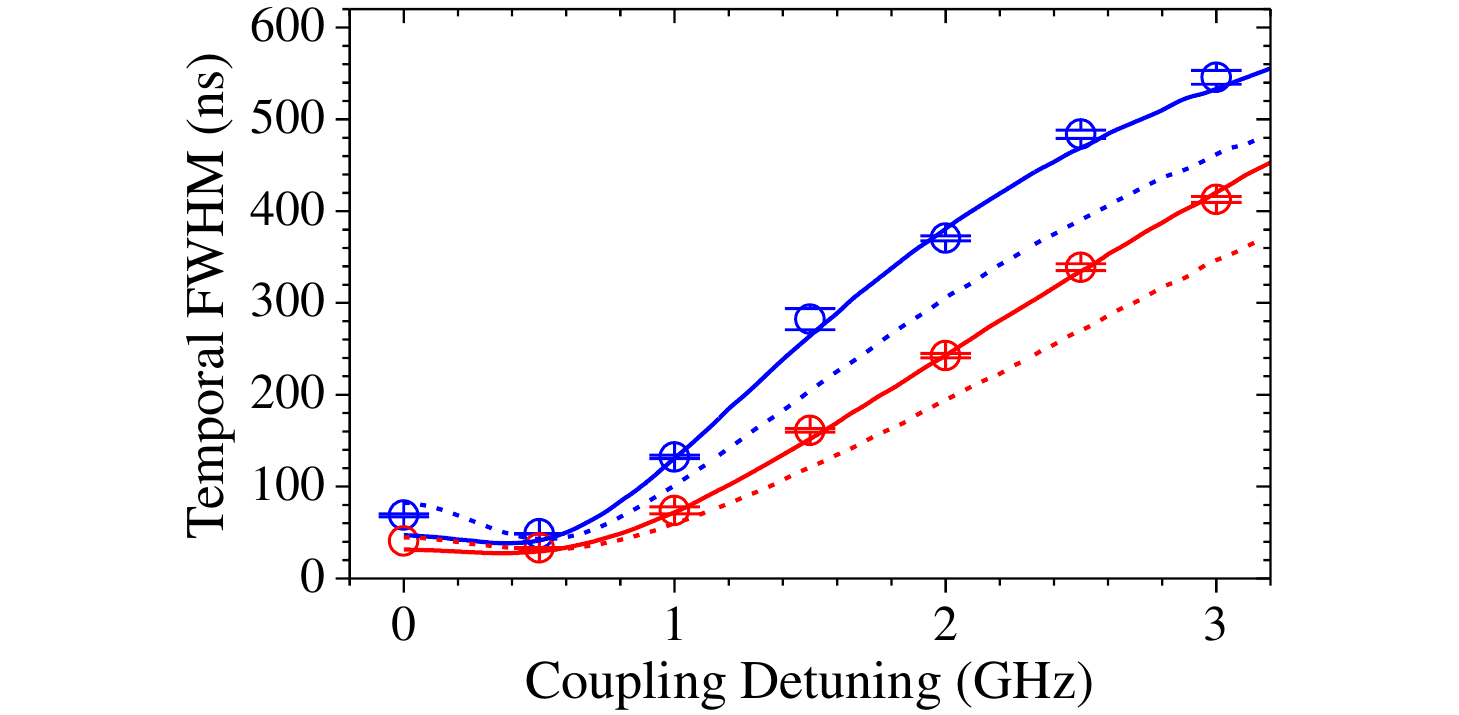}
	\caption{
Biphoton temporal FWHM as a function of the coupling detuning. The legends are identical to those in Fig.~\ref{fig:GR}. We obtain each temporal FWHM here and the corresponding generation rate in Fig.~\ref{fig:GR} (of the same color, symbol, and coupling detuning) from the same coincidence-count data. Solid and dashed lines (with and without the impurity atoms, respectively) here and those in Fig.~\ref{fig:GR} have the same calculation parameters.
	}
	\label{fig:TFWHM}
	\end{figure}
}


Biphotons are pairs of single photons. One can utilize one photon of a pair to herald the arrival of another in a quantum operation, so the former and latter are called the heralding and heralded single photons. Heralded single photons are practical and versatile qubits in long-distance quantum communication and optical quantum computation. The biphoton source based on the spontaneous four-wave mixing (SFWM) process with the double-$\Lambda$ transition scheme has the merits of narrow linewidth or long temporal width, high spectral brightness, stable frequency, and large linewidth tunability \cite{\CiteBiTheory}. The state-of-art double-$\Lambda$ biphoton sources with room-temperature or hot atoms have achieved a linewidth of 290~kHz (or 50~kHz with laser-cooled atoms) in Ref.~\cite{OurOPEX2021} (or Ref.~\cite{OurAPL2022}), a generation rate of 6.4$\times$10$^5$/s in this work, and a spectral brightness of 3.8$\times$10$^5$/s/MHz at a signal-to-background ratio (SBR) of 2.7 in Ref.~\cite{OurPRR2022} or 3.6$\times$10$^5$/s/MHz at a higher SBR of 6.8 in this work.

As shown in Fig.~\ref{fig:transition}, the double-$\Lambda$ scheme of the SFWM biphoton source consists of two Raman processes. The pump field and signal photons form the first Raman transition induced by the vacuum fluctuation, where population absorbs a pump photon, emits a signal photon, and moves from ground states $|1\rangle$ to $|2\rangle$. This generated signal photon moves out of the medium and typically becomes the heralding photon, and the Raman process establishes a ground-state coherence. The coupling field utilizes the ground-state coherence to induce the second Raman transition, where the population in $|2\rangle$ absorbs a coupling photon, emits a probe photon, and moves back to $|1\rangle$. This generated probe photon is typically the heralded photon. Each Ramann process maintains the two-photon resonance with the uncertainty due to the transition bandwidth. The two-photon resonance ensures the probe frequency follows the coupling frequency with the fixed offset of the frequency difference between $|1\rangle$ and $|2\rangle$. Therefore, one can vary the frequency of the coupling laser field to tune that of the heralded probe photons.

Previously in literature \cite{\CiteSfwmDl}, all of the double-$\Lambda$ SFWM biphoton sources employed resonant or near-resonant coupling laser fields to produce resonant or near-resonant probe single photons. The coupling-probe Raman transition with a nearly zero one-photon detuning exhibits the effect of electromagnetically induced transparency (EIT) or slow light, which prolongs the propagation delay time of probe photons and produces narrow-linewidth biphotons. A resonant probe field can experience the largest optical depth of a medium and possess the greatest propagation delay time due to the EIT effect. Furthermore, a resonant coupling field can enhance the nonlinear efficiency of the four-wave mixing process, and a larger coupling detuning results in a worse efficiency. More importantly, the previous theory of the double-$\Lambda$ SFWM biphoton generation reveals that the biphoton generation rate decreases with the coupling field detuning \cite{\CiteBiTheory}. Hence, in the double-$\Lambda$ SFWM biphoton source, using a far-detuned coupling field to generate far-detuned probe photons was considered counterintuitive.

Here, we discovered that the coupling field with a detuning, $\Delta_c$, far larger than the transition linewidth of a Doppler-broadened medium enhanced the biphoton generation rate.  At $\Delta_c$ = 0, the biphoton source had a generation rate, $R_g$, of 1.8$\times$$10^5$/s, a temporal full width at the half maximum (FWHM), $\tau_w$, of 64~ns, a spectral FWHM, $\Delta\omega/2\pi$, of 5.2~MHz, and a heralding probability, $h_p$, of 26\%. At $\Delta_c/2\pi$ = 1.0~GHz under the same experimental condition, the biphoton source not only enhanced $R_g$ to 6.4$\times$$10^5$/s but also improved $\tau_w$ to 132~ns, $\Delta\omega/2\pi$ to 1.8~MHz, and $h_p$ to 80\%. The previous theory of double-$\Lambda$ biphoton generation cannot predict the enhancement of $R_g$. We proposed an improved theory, whose physical origin will be discussed later. The proposed theory well explains the observed data. This study proposed and experimentally demonstrated a new knob for improving the performance of the double-$\Lambda$ SFWM biphoton source.

Furthermore, the proposed scheme makes both of the signal and probe photons far-detuned from their resonant frequencies in the atom-based biphoton source, similar to the signal and idler photons in the crystal- or chip-based biphoton source. In the past, the heralding efficiency or equivalently $h_p$ of the atom-based biphoton source was significantly lower than that of the crystal- or chip-based biphoton source \cite{\CiteSpdc}, resulting in a worse SBR due to unpaired probe or signal photons. Our data showed that the far-detuned coupling field and probe photons significantly increased $h_p$, or equivalently SBR, compared with the resonant ones under the same spectral brightness (SB) or generation rate per linewidth. For example, under a SB of 1.8$\times$$10^5$/s/MHz, the SBR was 2.5 at the resonant frequency and that was 12 at a far-detuned frequency. Tuning $\Delta_c$ enabled the SBR of the atom-based biphoton source to become commensurate with that of the crystal- or chip-based biphoton source under a similar SB.

\FigOne


We conducted the experiment with a vapor cell of isotopically enriched $^{87}$Rb atoms, heated to a temperature of approximately 48 $^\circ$C. In Fig.~\ref{fig:transition}, the states $|1\rangle$, $|2\rangle$, $|3\rangle$, and $|4\rangle$ represent $|5S_{1/2},F=2\rangle$, $|5S_{1/2},F=1\rangle$, $|5P_{1/2},F=1,2\rangle$, and $|5P_{3/2},F=1,2\rangle$, respectively. The frequency difference between the ground states $|1\rangle$ and $|2\rangle$ is approximately 6.8~GHz, and the spontaneous decay rates of the excited states $|3\rangle$ and $|4\rangle$ are all approximately 2$\pi\times$6~MHz.

The signal and probe photons were individually collected into two polarization-maintained optical fibers. The overall detection efficiencies ($D_s$ and $D_p$)  for the signal and probe photons, including the quantum efficiencies of the SPCMs, the peak transmittances of the etalons, and all optical losses, were 13$\pm1$\% and 9.4$\pm$0.5\%, respectively. We used only $D_s$ and $D_p$, excluding the collection efficiencies of the optical fibers, to derive $R_g$ and $h_p$ from the detection rate and the heralding efficiency. In this work, the quoted values $R_g$ and $h_p$ refer to the biphotons collected in the polarization-maintained optical fibers. Further details of the experimental setup can be found in \AppExpSetup.


We first explored the effects of $\Delta_c$ on the two-photon correlation function while keeping all other experimental parameters unchanged. Figure~\ref{fig:biphoton} shows the data of the two-photon correlation function (or number of the coincidence counts per time bin) versus the delay time, $\tau$, of the probe (heralded) photon's arrival relative to the detection or trigger of the signal (heralding) photon. The data also represent the biphoton wave packet. We fitted each biphoton wave packet with a phenomenological function \cite{OurPRR2022}. In each subfigure of Fig.~\ref{fig:biphoton}, the best fit matches the wave packet well. We performed the Fourier transform on the square root of the best-fit function and squared the result to obtain the biphoton spectrum as shown in each inset.

In each subfigure of Fig.~\ref{fig:biphoton}, we determined the SBR of biphotons from the best-fit curve. The right axis of the main plot shows the cross-correlation function between the signal and probe photons, $g^{(2)}_{s,p}(\tau)$. The SBR is approximately equal to the maximum value of $g^{(2)}_{s,p}$ subtracted by 1, where $g^{(2)}_{s,p}$ = 1 corresponds to the level of background counts. We used the area below the experimental data of the coincidence counts excluding the background counts and the accumulation time to calculate the detection rate, $R_d$. Due to the high photon counting rate, a saturation effect was existed. We removed the saturation effect from $R_d$ by using the formula in Eq.~(1) of Ref.~\cite{OurQST2025}. The corrected $R_d$ was divided by the product of $D_s$ and $D_p$ to obtain $R_g$. Not all the heralding events made by the signal photons resulted in heralded probe photons. We determined $h_p$ from the ratio of $R_g$ to the generation rate of single photons.

\FigTwo

Table~\ref{table:ForFigTwo} summarizes the results of Figs.~\ref{fig:biphoton}(a)-\ref{fig:biphoton}(d). It reveals the phenomenon that $R_g$ surprisingly increased with $\Delta_c$. The previous theory cannot predict such a phenomenon. We proposed an improved theory, which will be used to explain the phenomenon later. Table~\ref{table:ForFigTwo} also lists the values of $\tau_w$, $\Delta\omega$, SBR, and $h_p$. The behaviors of $\tau_w$ and $\Delta\omega$ versus $\Delta_c$ are expected from both the previous and proposed theories. We will also discuss the SBR and $h_p$ as functions of $\Delta_c$ later. Note that the quoted uncertainties in the table only take into account the measurement fluctuations.

\TableForFigTwo

Figure~\ref{fig:biphoton}(a) shows a biphoton wave packet at $\Delta_c$ = 0 as a reference for  Figs.~\ref{fig:biphoton}(b)-\ref{fig:biphoton}(d). When we increased $\Delta_c$, $\tau_w$ initially decreased and $R_g$ initially increased. The biphoton wave packet at $\Delta_c/2\pi$ = 0.2~GHz had the maximum SBR, a 1.3-time enhancement compared with the SBR at the resonance. As $\Delta_c/2\pi$ above 0.5~GHz approximately, $\tau_w$ started to increase and $R_g$ continuously increased.

When we increased $\Delta_c/2\pi$ to 0.7~GHz, the biphoton wave packet shown in Fig.~\ref{fig:biphoton}(c) had nearly the same $\tau_w$ and SBR as those in Fig.~\ref{fig:biphoton}(a). Compared with the resonance case, here $\Delta\omega/2\pi$ became 1.7-time narrower due to a slightly different temporal profile, and $R_g$ was 3.0-fold higher. In the past, a higher value of spectral brightness (SB) always resulted in a lower value of SBR by varying the OD, pump power and detuning, and/or coupling power \cite{OurPRR2022}. As we utilized tuning up $\Delta_c$ from 0 to 0.7~GHz, the biphoton wave packet's SB had a 5.2-time enhancement, while its SBR still maintained the same. This tuning knob of $\Delta_c$ has never been reported before.

In Fig.~\ref{fig:biphoton}(d), the biphoton wave packet at $\Delta_c/2\pi$ of 1.0~GHz increased $R_g$ by 3.6 folds, prolonged $\tau_w$ by 2.1 folds, narrowed down $\Delta\omega$ by 2.8 folds, and boosted the SB by 10 folds compared with the reference results. It also reached the maximum $R_g$ of 6.4$\times$10$^5$/s with respect to $\Delta_c$. Increasing $\Delta_c$ beyond 1.0~GHz monotonically decreased $R_g$, prolonged $\tau_w$, and narrowed down $\Delta\omega$.

From Figs.~\ref{fig:biphoton}(a) to \ref{fig:biphoton}(d), it is interesting to note the phenomenon that $h_p$ increased with $R_g$ as shown by Table~\ref{table:ForFigTwo}. The value of $h_p$ was linearly proportional to the value of $R_g$ for $\Delta_c/2\pi \leq$ 0.7~GHz, and it reached 80\% at $\Delta_c/2\pi$ = 1.0~GHz. The phenomenon inspired the proposed theory, which will be discussed in the next paragraph. Instead of using the signal photons to herald the probe photons, we also used the latter to herald the former \cite{OurQST2025} and increased $h_p$ to 86\% at $\Delta_c/2\pi$ = 1.0~GHz. This $h_p$ is the best value among all the double-$\Lambda$ biphoton sources.

\FigThree

\FigFour

As mentioned in the second paragraph, the Raman transition of the pump field and signal photons etstablishes the ground-state coherence, and the coupling field utilizes the coherence to generate the probe photons. In a Doppler-broadened medium, due to the thermal motion atoms in the beam path of the probe photons might not carry this ground-state coherence or might carry a coherence but with a mismatched phase. Such atoms are more prominent in room-temperature or hot media, and they become impurities in the biphoton generation process. Each impurity atom behaves like a two-level system of the transition from $|1\rangle$ to $|3\rangle$ that can absorb a heralded probe photon. Nevertheless, a far-detuned frequency can protect the probe photons from being absorbed by the impurity atoms. With the resonant frequency, the ratio of the value of $h_p$ with the probe photons heralding the signal photons to that with the latter heralding the former was 2.2 (or 3.9) as demonstrated by Table~\ref{table:ForFigTwo} (or by Ref.~\cite{OurQST2025}). With a far-detuned frequency, the ratio of the two values of $h_p$ was 1.08 as demonstrated by the case of 1.0~GHz in Table~\ref{table:ForFigTwo}. The significant difference between the two ratios is an evidence for the protection of the probe photons by a far-detuned frequency. The above argument explains the physical origin of the observed data of $R_g$ in Fig.~\ref{fig:biphoton}.  We took into account the impurity atoms and derived the two-photon correlation function $G^{(2)}(\tau)$ shown in \AppTheory. The parameters $b$, $\Omega_c$, and $\gamma$ in $G^{(2)}(\tau)$ represent the fraction of impurity atoms in the medium, coupling Rabi frequency, and ground-state decoherence rate. 

We next changed the pump power ($P_p$) and coupling power ($P_c$) and measured $R_g$ and $\tau_w$ as functions of $\Delta_c$ as shown in Figs.~\ref{fig:GR} and \ref{fig:TFWHM}, respectively. The blue circles in the fgures were measured with a similar $P_c$ but a significantly different $P_p$ compared with the data in  Fig.~\ref{fig:biphoton}. The red circles in the figures were measured with the same $P_p$ but a very different $P_c$ compared with the blue circles. All of the data provide evidence that $P_c$ and $P_p$ have very little effect on the dependence of $R_g$ on $\Delta_c$. The evidence is consistent with the proposed theory that each impurity atom behaves like a two-level system. 

To test the proposed theory, we simultaneously fit the blue (and red) circles in Figs.~\ref{fig:GR} and \ref{fig:TFWHM} with the theoretical predictions from the formulas of $G^{(2)}(\tau)$ and $R_g$ in \AppTheory~by varying $b$, $\Omega_c$, and $\gamma$. The blue and red solid lines are the best fits. They give the similar values of $b$ (also $\gamma$), and provide the values of $\Omega_c^2$ approximately linearly proportional to those of $P_c$. Hence, $G^{(2)}(\tau)$ and $R_g$ of the proposed theory can satisfactorily predict the data. On the contrary, the dashed lines calculated from the previous theory are completely incompatible with the data of $R_g$.

The green squares in Fig.~\ref{fig:GR} represent the generation rates of the signal photons, which varied slightly with $\Delta_c$. The ratio of each data point of $R_g$ to the corresponding signal-photon generation rate gives $h_p$. The increment of $h_p$ by $\Delta_c$ is consistent with the protection of the probe photons and the enhancement of probe transmittance. As $\Delta_c/2\pi$ became very large, the protection effect saturated or the probe transmittance was nearly unity. The decrement of $h_p$ in the regime of $\Delta_c/2\pi$ beyond 1.5~GHz is due to the degradation of FWM efficiency.

The increment of $h_p$ indicates that more signal and probe photons become paired, and fewer unpaired photons contribute to the background, which enhances the SBR. Furthermore, the fluorescence photons induced by the resonant coupling field contributed up to about 40\% of the counts in the probe detection channel \cite{OurQST2025}, and they were the major source of the background photons in the past \cite{OurPRR2022, OurOPEX2024}. Using a far-detuned $\Delta_c$ can significantly reduce the background and improve the SBR. Finally, in this work, we compared the SB and SBR at the resonant frequency with those at the far-detuned frequencies. Under a similar SBR of 6.8 or 6.6, the SB of $\Delta_c/2\pi$ = 1.0~GHz is 4.6-fold better than that of $\Delta_c$ = 0. Under the same SB of 1.8$\times$$10^5$/s/MHz, the SBR of $\Delta_c/2\pi$ = 0.7~GHz is 4.8-fold better than that of $\Delta_c$ = 0. Experimental data are shown in \AppSBR. The comparison demonstrates that tuning $\Delta_c$ is an effective method to improve the SBR of a high-SB double-$\Lambda$ biphoton source.


In conclusion, we proposed and experimentally demonstrated tuning $\Delta_c$ to protect the heralded probe photons from being absorbed by the impurity atoms. Tuning $\Delta_c$ far away from the resonance was counter-intuitive according the previous theory of double-$\Lambda$ SFWM biphoton generation. Here, the biphoton wave packet at $\Delta_c/2\pi$ = 1.0~GHz increased $R_g$ by 3.6 folds, $\tau_w$ by 2.1 folds, the SB by 10 folds, and $h_p$ by 3.0 folds compared with that at $\Delta_c$ = 0 under the same experimental condition. Previously, varying the OD, pump power, pump detuning, and/or coupling power to achieve a higher value of SB always results in a lower value of SBR. Here, the biphoton wave packet at $\Delta_c/2\pi$ = 0.7~GHz enhanced the SB by 5.2 folds and maintained the same SBR compared with that at $\Delta_c$ = 0 under the same experimental condition. This work opens a new avenue for enhancing the performance of the double-$\Lambda$ SFWM biphoton source, and makes the SBR of the source become comparable with that of the crystal- or chip-based biphoton source under a similar SB.

\section*{ACKNOWLEDGMENTS}

This work was supported by Grants No.~112-2112-M-007-020-MY3 and No.~113-2119-M-007-012 of the National Science and Technology Council, Taiwan.


\newpage \newpage

{\centering{\large \bf SUPPLEMENTAL MATERIALS \\}} \vspace*{-\baselineskip}

\newcommand{\MainFigOne}{Fig.~1}
\newcommand{\MainFigTwo}{Fig.~2}
\newcommand{\MainFigTwoA}{Fig.~2(a)}
\newcommand{\MainFigTwoC}{Fig.~2(c)}
\newcommand{\MainFigTwoD}{Fig.~2(d)}
\newcommand{\MainTableI}{Table~I}

\renewcommand{\theequation}{S\arabic{equation}}
\setcounter{equation}{0}
\renewcommand{\thefigure}{S\arabic{figure}}
\setcounter{figure}{0}
\renewcommand{\thetable}{S\arabic{figure}}
\setcounter{table}{0}

\newcommand{\FigSBR}{
	\begin{figure}[t]
	\includegraphics[width=\columnwidth]{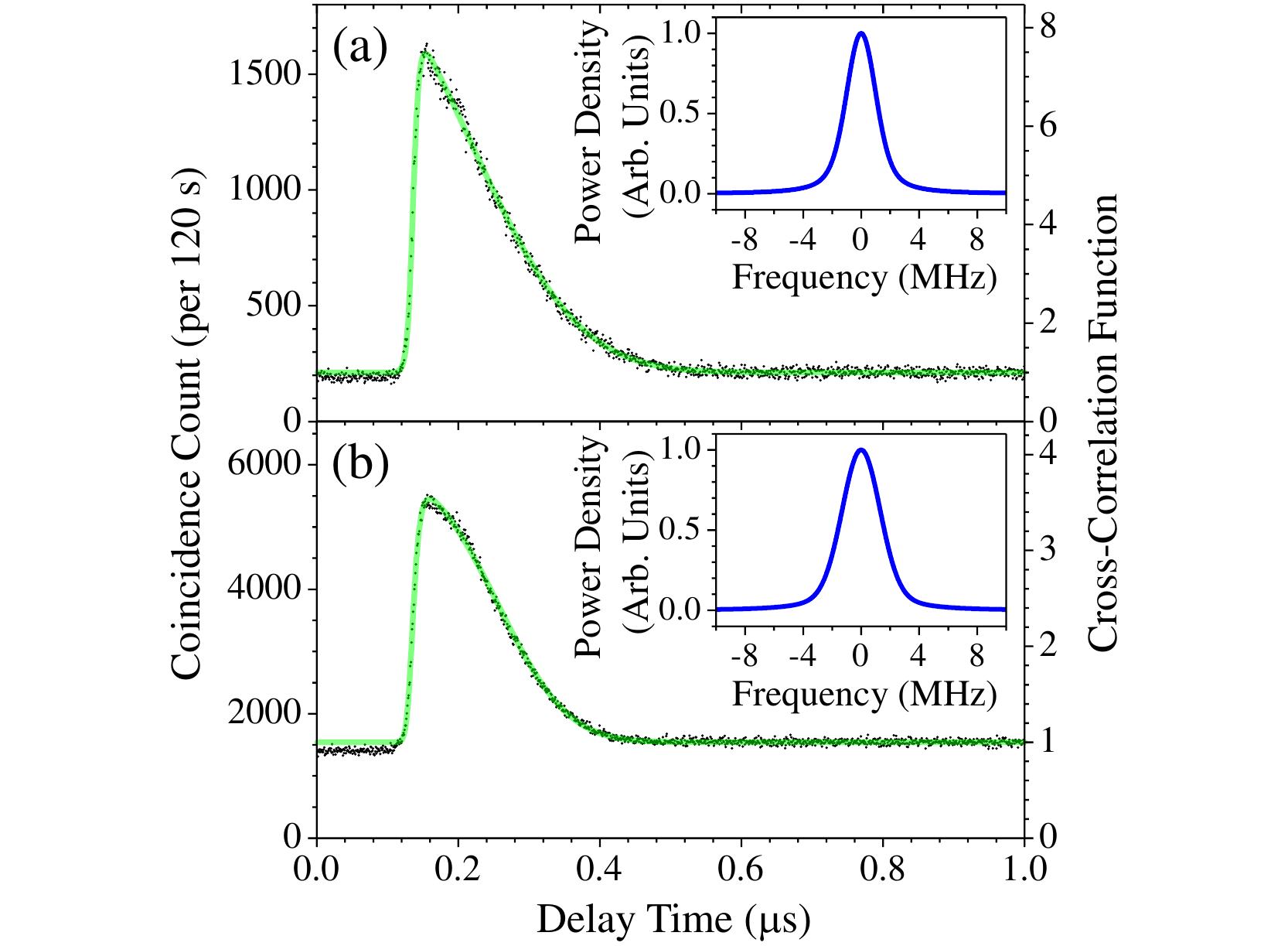}
	\caption{
Data showing coincidence count (left axis) and cross-correlation (right axis) versus the delay time between the signal and probe photons. Black dots represent the experimental data, and green lines are the best fits. Each inset shows the frequency spectrum of the best fit. The OD and pump detuning were the same as those in \MainFigTwo~of the main text. We set $\Delta_c$ to zero and varied the pump power ($P_p$) and coupling power ($P_c$). (a) $P_p$ = 5~mW and $P_c$ = 7~mW, resutling in $R_g$ = 1.9$\times10^5$/s, $\Delta\omega/2\pi$ = 2.5~MHz, and SBR = 6.6. (b) $P_p$ = 15~mW and $P_c$ = 6~mW, resutling in $R_g$ = 5.8$\times10^5$/s, $\Delta\omega/2\pi$ = 3.2~MHz, and SBR = 2.5.
	}
	\label{fig:biphotonSBR}
	\end{figure}
}
\newcommand{\TableForFigSBR}{
	\begin{table}[t]
	\caption{Compare the biphotons' SB and SBR of $\Delta_c$ = 0 with those of $\Delta_c$ = 0.7 and 1.0~GHz.
	}
{\centering
	\footnotesize
	\begin{tabular}{c c c c} 
	\hline 
	\parbox{15mm}{Figure\\number} &
	\parbox{15mm}{\vspace*{1mm}
		$\Delta_c/2\pi$\\ \vspace*{2pt}(GHz)
				\vspace*{1mm}} &
	\parbox{28mm}{SB$^\ast$\\ \vspace*{2pt}($10^5$/s/MHz)} &
	\parbox{15mm}{SBR} \\
 	\hline 
	\ref{fig:biphotonSBR}(a) & 
		0.0 & 0.76$\pm$0.02 & 6.57$\pm$0.03 \\
	2(d) of main text & 
		1.0 & 3.51$\pm$0.07 & 6.8$\pm$0.1 \\
	\hline
	\ref{fig:biphotonSBR}(b) & 
		0.0 & 1.80$\pm$0.04 & 2.5$\pm$0.1 \\
	2(c) of main text & 
		0.7 & 1.77$\pm$0.05 & 11.9$\pm$0.3 \\ 
	\hline
	\end{tabular}
	\hspace*{3.0pt}
	\parbox{87mm}{
		\vspace*{-2mm}
		\begin{flushleft}
		\hspace*{-5.0pt}$^{\ast}${These values of SB refer to the biphotons 
			collected in the polarization-maintained optical fibers. Multiply each 
			value by a factor of 1.9 to obtain the SB right after the vapor cell.}\\
		\end{flushleft}
	}
}
	\label{table:ForFigSBR}
	\end{table}
}

\section{Experimental Setup}

We utilized the energy levels of $|5S_{1/2},F=2\rangle$, $|5S_{1/2},F=1\rangle$, $|5P_{1/2},F=1,2\rangle$, and $|5P_{3/2},F=1,2\rangle$ in a biphoton generation experiment of $^{87}$Rb atomic vapor heated to 48 $^\circ$C. These states are denoted by $|1\rangle$, $|2\rangle$, $|3\rangle$, and $|4\rangle$, respectively. The pump laser field drove the transition from $|1\rangle$ to $|4\rangle$. Its detuning, $\Delta_{p}$, refers to the transition from $|1\rangle$ to $|5P_{3/2},F=2\rangle$. The coupling laser field drove the transition from $|2\rangle$ to $|3\rangle$. Its detuning, $\Delta_c$, refers to the transition from $|2\rangle$ to $|5P_{1/2},F=2\rangle$. The pump field and signal photon formed the first two-photon Raman transition of $|1\rangle$ $\rightarrow$ $|4\rangle$ $\rightarrow$ $|2\rangle$. The coupling field and probe photon formed the second two-photon Raman transition of $|2\rangle$ $\rightarrow$ $|3\rangle$ $\rightarrow$ $|1\rangle$. These two Raman transitions compose the double-$\Lambda$ spontaneous four-wave mixing (SFWM) process. The spontaneous decay rates of the excited states $|3\rangle$ and $|4\rangle$ are all about $2\pi\times$6~MHz denoted as $\Gamma$.

In the experiment, the pump and coupling fields and the signal and probe photons propagated in the same direction and completely overlapped inside the atomic vapor cell. This all-copropagation scheme minimizes the phase mismatch of the FWM process as well as the decay rate of the ground-state coherence \cite{OurOPEX2021, OurPRR2022}. We further applied a hyperfine optical pumping (HOP) field to place all the population in $|1\rangle$. The HOP field has a hollow core centered around the region of biphoton generation. In the presence of the laser fields in the biphoton generation, we used the absorption spectrum of a weak probe laser beam to determine the optical depth (OD) of the medium used in the Eqs.~(\ref{eq:FWM})-(\ref{eq:EITm}).

The pump and coupling laser fields had the $p$ and $s$ polarizations, and the signal and probe photons had the $s$ and $p$ polarizations. We installed the polarization filters and etalons to block the two laser fields, where the total extinction ratios of the laser beams were higher than 140~dB. Two single-photon counting module (SPCM) of Excelitas SPCM-AQRH-13-FC detected the signal and probe photons. Each SPCM had a dark count rate of about 200 counts/s. The total of the pump (or coupling) laser leakage and the signal (or probe) SPCM's dark count had a negligible contribution, i.e., approximately $< 3$\%, to the signal (or probe) photon detection. We collected the signal and probe photons individually into two polarization-maintained optical fibers, which had collection efficiencies ($C_s$ and $C_p$) of 75\% and 69\% for the spatial modes of the pump and coupling laser fields, respectively. The overall detection efficiencies ($D_s$ and $D_p$), including the SPCMs' quantum efficiencies, etalons' peak transmittances, and all optical losses but excluding $C_s$ and $C_p$, were 13$\pm1$\% and 9.4$\pm$0.5\% for the signal and probe photons. We used only $D_s$ and $D_p$ without $C_s$ and $C_p$ to derive the generation rate and heralded probability from the detection rate and heralding efficiency. Other details of the experimental setup can be found in Ref.~\cite{OurQST2025}.

\section{Two-Photon Correlation Function}

In the main text, we have discussed the physical origin of the phenomenon that the generation rate of the double-$\Lambda$ SFWM biphoton source increases with coupling detuning. A theoretical model based on its physical origin gives the following formula for the biphoton wave packet or two-photon correlation function, $G^{(2)}(\tau)$, of a Doppler-broadened medium.
\begin{eqnarray}
\label{eq:biphoton}
	\hspace*{-3.5pt}
	G^{(2)}(\tau) \hspace*{-5pt} && =
		\left| 
		\int_{-\infty}^{\infty} d\delta \frac{e^{-i\delta\tau}}{2\pi}
		\bar{\kappa}(\delta)\,
		{\rm sinc} [\bar{\rho}_c(\delta) + \bar{\rho}_m(\delta)]
		\right. \nonumber \\
		&& \times
		\left. 
		e^{i[\bar{\rho_c}(\delta) + \bar{\rho}_m(\delta)]}
		B(\delta)
		\right|^2,
\end{eqnarray}
where $\delta$ is the two-photon detuning, $\bar{\kappa}(\delta)$ is proportional to the cross-susceptibility between the signal and probe photons, $\bar{\rho}_c(\delta)$ and $\bar{\rho}_m(\delta)$ are proportional to the self-susceptibilities of the probe photons induced by the atoms with the ground-state coherence and the impurity atoms, respectively, and $B(\delta)$ describes the combined frequency spectrum of the probe and signal etalons. The four functions of $\delta$ in the integral are defined as follows:
\begin{eqnarray}
\label{eq:FWM}
	\bar{\kappa}(\delta) \hspace*{-5pt} && =
		\frac{(1-b)\alpha}{4}\int_{-\infty}^{\infty} d\omega_D
		\left[  
		\frac{e^{-\omega_D^2/\Gamma_D^2}}{\sqrt{\pi}\Gamma_D}
		\frac{\Omega_p}{\Delta_p + \omega_D + i\Gamma/2}
		\right.
		\nonumber \\
		&& \times
		\left.
		\frac{\Omega_c\Gamma}{\Omega_c^2-4(\delta+i\gamma)
		(\delta+\Delta_c+\omega_D+i\Gamma/2)}
		\right],
\end{eqnarray}
\begin{eqnarray}
\label{eq:EITc}
	\bar{\rho}_c(\delta) \hspace*{-5pt} && =
		\frac{(1-b)\alpha}{2} \int_{-\infty}^{\infty} d\omega_D
		\left[
		\frac{e^{-\omega_D^2/\Gamma_D^2}}{\sqrt{\pi}\Gamma_D}
		\right.
		\nonumber \\
		&& \times
		\left.
		\frac{(\delta+i\gamma)\Gamma}
		{\Omega_c^2-4(\delta+i\gamma)(\delta+\Delta_c+\omega_D+i\Gamma/2)}
		\right],
		\\
\label{eq:EITm}
	\bar{\rho}_m(\delta) \hspace*{-5pt} && =
		\frac{b\alpha}{2} \int_{-\infty}^{\infty} d\omega_D
		\left[
		\frac{e^{-\omega_D^2/\Gamma_D^2}}{\sqrt{\pi}\Gamma_D}
		\right.
		\nonumber \\
		&& \times
		\left.
		\frac{\Gamma}
		{4(\delta+\Delta_c+\omega_D+i\Gamma/2)}
		\right],
		\\
\label{eq:etalon}
	B(\delta) \hspace*{-5pt} && =
		\left( \frac{1}{1+4\delta^2/\Gamma_e^2} \right)^2,
\end{eqnarray} 
where $\omega_D$ represents the Doppler shift due to atomic velocity, $\alpha$ is the optical depth of the medium, $b$ is the fraction of the impurity atoms, $\Gamma$ ($\approx 2\pi\times$6~MHz) is the spontaneous decay rate of the excited states, $\Omega_p$ and $\Omega_c$ denote the Rabi frequencies of the pump and coupling fields, $\Delta_p$ and $\Delta_c$ are the pump's and coupling's one-photon detunings, $\gamma$ represents the decoherence rate in the experimental system, and $\Gamma_e$ is the effective width of the etalons.

\FigSBR

\TableForFigSBR

The biphoton generation rate, $R_g$, is proportional to the integral of the wave packet over the delay time given by
\begin{equation}
	R_g  \propto \int_{-\infty}^{\infty} d\tau G^{(2)}(\tau).
\end{equation} 

In the theoretical calculation, we set $\Gamma_D = 54\Gamma$ based on the vapor cell's temperature, $\alpha = 500$ which was determined from the absorption spectrum, and $\Gamma_e$ = 53.5~MHz or 8.9$\Gamma$ which was determined by sweeping the laser frequencies to measure the etalons' transmission spectra. We varied $b$, $\Omega_c$, and $\gamma$ to fit the experimental data of $\tau_w$ and $R_g$ simultaneously with the theoretical predictions.

\vspace*{\baselineskip}

\section{Spectral Brightness and Signal-to-Background Ratio}

We attempted to make the biphoton wave packet at the resonant frequency or $\Delta_c$ = 0 perform better by varying the pump and coupling powers, $P_p$ and $P_c$, respectively. Considering the double-$\Lambda$ SFWM process, increasing $P_p$ enlarges the generation rate with approximately the linear dependence and does not affect the biphoton temproal profile and width, and decreasing $P_c$ prolongs the temporal width significantly and changes the generation a little. Note that varying $P_p$ and $P_c$ to achieve a higher value of spectral brightness (SB) or generation rate per linewidth always results in a lower value of signal-to-background ratio (SBR) \cite{OurOPEX2021, OurPRR2022}. Given a SBR (or a SB), we searched for the best SB (or SBR) at $\Delta_c$ = 0.

The biphoton wave packet at $\Delta_c/2\pi$ = 1.0~GHz with $P_p$ of 5~mW and $P_c$ of 17~mW achieved the highest SB of 3.5$\times$10$^5$/s/MHz in this work. It also had a SBR of 6.8. In the main text, \MainFigTwoD~and \MainTableI~show the data and results. To get a better SB, one can employ the methods of increasing $P_p$ to enhance the generation rate, decreasing $P_c$ to narrow down the spectral linewidth, or both. All of the methods resulted in a smaller SB at a SBR around 6.8. Figure~\ref{fig:biphotonSBR}(a) shows the representative biphoton wave packet with a SBR of 6.6 at $\Delta_c$ = 0. However, its SB is still 4.6 times worse than that in \MainFigTwoD. Please see the comparison in Table~\ref{table:ForFigSBR}. Note that the quoted uncertainties in the table only take into account the measurement fluctuations.

We further test whether, at a given SB, the advantage in SBR of a detuned frequency is also significant. The biphoton wave packet at $\Delta_c/2\pi$ = 0.7~GHz with $P_p$ of 5~mW and $P_c$ of 17~mW achieved a SB of 1.8$\times$10$^5$/s/MHz and a SBR of 12. In the main text, \MainFigTwoC~and \MainTableI~show the data and results. To obtain the same SB, we increased $P_p$ by 3 folds to enhance the generation rate and decreased $P_c$ by about 2.8 folds to narrow down the biphoton linewidth. Figure~\ref{fig:biphotonSBR}(b) shows the biphoton wave packet with a SB of 1.8$\times$10$^5$/s/MHz at $\Delta_c$ = 0. However, its SBR is 4.8 times worse than the SBR in \MainFigTwoC. Please see the comparison in Table~\ref{table:ForFigSBR}. The advantage in SBR of $\Delta_c$ = 0.7~GHz is significant compared with the resonant case.

\begin{thebibliography}{99}

\bibitem{BiphotonTheory.2008}
	S. Du, J. Wen, and M. H. Rubin, 
	\RT{Narrowband biphoton generation near atomic resonance} 
	J. Opt. Soc. Am. B {\bf 25}, C98 (2008).
\bibitem{OurAQT2024}
	J.-M. Chen, T. Peters, P.-H. Hsieh, and I. A. Yu,
	\RT{Review of Biphoton Sources based on the Double-$\Lambda$ Spontaneous Four-Wave Mixing Process} 
	Adv. Quantum Technol. {\bf 7}, 2400138 (2024).

\bibitem{OurOPEX2021}
	C.-Y. Hsu, Y.-S. Wang, J.-M. Chen, F.-C. Huang, Y.-T. Ke, E. K. Huang, W. Hung, K.-L. Chao, S.-S. Hsiao, Y.-H. Chen, C.-S. Chuu, Y.-C. Chen, Y.-F. Chen, I. A. Yu,
	\RT{Generation of sub-MHz and spectrally-bright biphotons from hot atomic vapors with a phase mismatch-free scheme}
	Opt. Express {\bf 29}, 4632 (2021).
\bibitem{OurAPL2022}
	Y.-S. Wang, K.-B. Li, C.-F. Chang, T.-W. Lin, J.-Q. Li, S.-S. Hsiao, J.-M. Chen, Y.-H. Lai, Y.-C. Chen, Y.-F. Chen, C.-S. Chuu, and I. A. Yu, 
	\RT{Temporally ultralong biphotons with a linewidth of 50 kHz}
	APL Photonics {\bf 7}, 126102 (2022).
\bibitem{OurPRR2022}
	J.-M. Chen, C.-Y. Hsu, W.-K. Huang, S.-S. Hsiao, F.-C. Huang, Y.-H. Chen, C.-S. Chuu, Y.-C. Chen, Y.-F. Chen, and I. A. Yu, 
	\RT{Room-temperature biphoton source with a spectral brightness near the ultimate limit}
	Phys. Rev. Res. {\bf 4}, 023132 (2022).

\bibitem{SFWM.DL.LCA.2005}
	V. Bali\'{c}, D. A. Braje, P. Kolchin, G. Y. Yin, and S. E. Harris,
	\RT{Generation of Paired Photons with Controllable Waveforms,}
	Phys. Rev. Lett. {\bf 94}, 183601 (2005).
\bibitem{SFWM.DL.LCA.2006}
    J. K. Thompson, J. Simon, H. Loh, and V. Vuleti\'{c},
	\RT{A high-brightness source of narrowband, identical-photon pairs}
	Science {\bf 313}, 74-76 (2006).
\bibitem{SFWM.DL.LCA.2008}
	P. Kolchin, C. Belthangady, S. Du, G. Y. Yin, and S. E. Harris,
	\RT{Electro-Optic Modulation of Single Photons}
	Phys. Rev. Lett. {\bf 101}, 103601 (2008).
\bibitem{SFWM.DL.LCA.2009}
	C. Belthangady, S. Du, C.-S. Chuu, G. Y. Yin, and S. E. Harris,
	\RT{Modulation and measurement of time-energy entangled photons}
	Phys. Rev. A {\bf 80}, 031803 (2009).
\bibitem{SFWM.DL.LCA.2010}
	S. Yun, J. Wen, P. Xu, M. Xiao, and S.-N. Zhu,
	\RT{Generation of frequency-correlated narrowband biphotons from four-wave mixing in cold atoms}
	Phys. Rev. A {\bf 82}, 063830 (2010).
\bibitem{SFWM.DL.LCA.2011}
	H. Yan , S. Zhang, J. F. Chen, M. M. T. Loy, G. K. L. Wong, and S. Du,
	\RT{Generation of Narrow-Band Hyperentangled Nondegenerate paired photons}
	Phys. Rev. Lett. {\bf 106}, 033601 (2011).
\bibitem{SFWM.DL.LCA.2014a}
	L. Zhao, X. Guo, C. Liu, Y. Sun, M. M. T. Loy, and S. Du,
	\RT{Photon pairs with coherence time exceeding 1 $\mu$s}
	Optica {\bf 1}, 84-88 (2014).
\bibitem{SFWM.DL.LCA.2014b}
	C. Liu, Y. Sun, L. Zhao, S. Zhang, M. M. T. Loy, and S. Du,
	\RT{Efficiently loading a single photon into a single-sided fabry-perot cavity}
	Phys. Rev. Lett. {\bf 113}, 133601 (2014).
\bibitem{SFWM.DL.LCA.2015a}
	Z. Han, P. Qian, L. Zhou, J. F. Chen, and W. Zhang,
	\RT{Coherence time limit of the biphotons generated in a dense cold atom cloud}
	Sci. Rep. {\bf 5}, 9126 (2015).
\bibitem{SFWM.DL.LCA.2015b}
    L. Zhao, X. Guo, Y. Sun, Y. Su, M. M. T. Loy, and S. Du,
	\RT{Shaping the biphoton temporal waveform with spatial light modulation}
	Phys. Rev. Lett. {\bf 115}, 193601 (2015).
\bibitem{SFWM.DL.LCA.2016a}
	L. Zhao, Y. Su, and S. Du,
	\RT{Narrowband biphoton generation in the group delay regime}
	Phys. Rev. A {\bf 93}, 033815 (2016).
\bibitem{SFWM.DL.LCA.2016b}
	P. Farrera, G. Heinze, B. Albrecht, M. Ho, M. Ch\'{a}vez, C. Teo, N. Sangouard, and H. de Riedmatten,
	\RT{Generation of single photons with highly tunable wave shape from a cold atomic ensemble}
	Nat. Commun. {\bf 7}, 13556 (2016).
\bibitem{SFWM.DL.LCA.2018}
	C. Yang, Z. Gu, P. Chen, Z. Qin, J. F. Chen, and W. Zhang,
	\RT{Tomography of the Temporal-Spectral State of Subnatural-Linewidth Single Photons from Atomic Ensembles}
	Phys. Rev. Applied {\bf 10}, 054011 (2018).
\bibitem{SFWM.DL.LCA.2019a}
    Y. Wang, J. Li, S. Zhang, K. Su, Y. Zhou, K. Liao, S. Du, H. Yan, and S.-L Zhu,
	\RT{Efficient quantum memory for single-photon polarization qubits}
	Nat. Photonics {\bf 13}, 346 (2019).
\bibitem{SFWM.DL.LCA.2019b}
    J.-F. Li, Y.-F. Wang, K.-Y. Su, K.-Y. Liao, S.-C. Zhang, H. Yan, and S.-L. Zhu,
	\RT{Generation of Gaussian-shape single photons for high efficiency quantum storage}
	Chinese Phys. Lett. {\bf 36}, 074202 (2019).
\bibitem{SFWM.DL.LCA.2020}
	R. Chinnarasu, C.-Y. Liu, Y.-F. Ding, C.-Y. Lee, T.-H. Hsieh, I. A. Yu, and C.-S. Chuu,
	\RT{Efficient generation of subnatural-linewidth biphotons by controlled quantum interference}
	Phys. Rev. A {\bf 101}, 063837 (2020).
\bibitem{SFWM.DL.LCA.2023}
    A. Bruns, C.-Y. Hsu, S. Stryzhenko, E. Giese, L. P. Yatsenko, I. A. Yu, T. Halfmann, and T. Peters,
	\RT{Ultrabright and narrowband intra-fiber biphoton source at ultralow pump power}
	Quantum Sci. Technol. {\bf 8}, 015002 (2023).
\bibitem{YFCPRR2024}
	J.-S. Shiu, Z.-Y. Liu, C.-Y. Cheng, Y.-C. Huang, I. A. Yu, Y.-C. Chen, C.-S. Chuu, C.-M. Li, S.-Y. Wang, and Y.-F. Chen,
	\RT{Observation of highly correlated ultrabright biphotons through increased atomic ensemble density in spontaneous four-wave mixing}
	Phys. Rev. Res. {\bf 6}, L032001 (2024).

\bibitem{SFWM.DL.RHA.2016}
	C. Shu, P. Chen, T. K. A. Chow, L. Zhu, Y. Xiao, M. M. T. Loy, and S. Du,
	\RT{Subnatural-linewidth biphotons from a Doppler-broadened hot atomic vapour cell}
	Nat. Commun. {\bf 7}, 12783 (2016).
\bibitem{SFWM.DL.RHA.2017a}
	L. Zhu, X. Guo, C. Shu, H. Jeong, and S. Du,
	\RT{Bright narrowband biphoton generation from a hot rubidium atomic vapor cell}
	Appl. Phys. Lett. {\bf 110}, 161101 (2017).
\bibitem{SFWM.DL.RHA.2017b}
    L. Podhora, P. Ob\v{s}il, I. Straka, M. Je\v{z}ek, and L. Slodi\v{c}ka,
	\RT{Nonclassical photon pairs from warm atomic vapor using a single driving laser}
	Opt. Express {\bf 25}, 31230 (2017).
\bibitem{SFWM.DL.RHA.2018}
	X. Li, D. Zhang, D. Zhang, L. Hao, H. Chen, Z. Wang, and Y. Zhang,
	\RT{Dressing control of biphoton waveform transitions}
	Phys. Rev. A {\bf 97}, 053830 (2018).
\bibitem{SFWM.DL.RHA.2019}
	Y. Liu, K. Li, S. Zhang, H. Fan, W. Li, and Y. Zhang, 
	\RT{Generation of correlated biphoton via four wave mixing coexisting with multi-order fluorescence processes}
	Sci. Rep. {\bf 9}, 20065 (2019).
\bibitem{SFWM.DL.RHA.2020a}
    J. Mika, and L. Slodi\v{c}ka,
	\RT{High nonclassical correlations of large-bandwidth photon pairs generated in warm atomic vapor}
	J. Phys. B: At. Mol. Opt. Phys. {\bf 53}, 145501 (2020).
\bibitem{SFWM.DL.RHA.2020b}
	T. Jeong and H. S. Moon,
	\RT{Temporal- and spectral-property measurements of narrowband photon pairs from warm double-$\Lambda$-type atomic ensemble}
	Opt. Express {\bf 28}, 3985-3994 (2020).
\bibitem{SFWM.DL.RHA.2021}
	K. Li, Y. Zhao,Y. Qin, Z. Chen, Y. Cai, and Y. Zhang,
	\RT{Shaping Temporal Correlation of Biphotons in a Hot Atomic Ensemble}
	Adv. Photonics Res. {\bf 2}, 2100073 (2021).
\bibitem{SFWM.DL.RHA.2022}
    J. Mika, L. Lachman, T. Lamich, R. Filip, and L. Slodi\v{c}ka,
	\RT{Single-mode quantum non-Gaussian light from warm atoms}
	npj Quantum Inf. {\bf 8}, 123 (2022).
\bibitem{OurPRA2022}
	S.-S. Hsiao, W.-K. Huang, Y.-M. Lin, J.-M. Chen, C.-Y. Hsu, and I. A. Yu, 
	\RT{Temporal profile of biphotons generated from a hot atomic vapor and spectrum of electromagnetically induced transparency} 
	Phys. Rev. A {\bf 106}, 023709 (2022).
\bibitem{OurOPEX2024}
	T.-J. Shih, W.-K. Huang, Y.-M. Lin, K.-B. Li, C.-Y. Hsu, J.-M. Chen, P.-Y. Tu, T. Peters, Y.-F. Chen, and I. A. Yu,
	\RT{A universal relation between conditional auto-correlation function and cross-correlation function of biphotons} 
	Opt. Express {\bf 32}, 13657 (2024).
\bibitem{OurQST2025}
	W.-K. Huang, B. Kim, T.-J. Shih, C.-Y. Hsu, P.-Y. Tu, T.-Y. Lin, Y.-F. Chen, C.-S. Chuu, and I. A. Yu,
	\RT{Time-reversing biphoton source of the double-$\Lambda$ spontaneous four-wave mixing process} 
	accepted for the publication in Quantum Sci. Technol. https://doi.org/10.1088/2058-9565/ada08f. 

\bibitem{SPDC.2012}
	C.-S. Chuu, G. Y. Yin, and S. E. Harris,
	\RT{A miniature ultrabright source of temporally long, narrowband biphotons}
	Appl. Phys. Lett. {\bf 101}, 051108 (2012). 
\bibitem{SPDC.2013}
	J. Fekete, D. Riel{\"{a}}nder, M. Cristiani, and H. de Riedmatten,
	\RT{Ultranarrow-Band Photon-Pair Source Compatible with Solid State Quantum Memories and Telecommunication Networks}
	Phys. Rev. Lett. {\bf 110}, 220502 (2013).
\bibitem{SPDC.2015a}
	A. Lenhard, M. Bock, C. Becher, S. Kucera, J. Brito, P. Eich, P. M{\"{u}}ller, and J. Eschner,
	\RT{Telecom-heralded single-photon absorption by a single atom}
	Phys. Rev. A {\bf 92}, 063827 (2015).
\bibitem{SPDC.2015b}
	K.-H. Luo, H. Herrmann, S. Krapick, B. Brecht, R. Ricken, V. Quiring, H. Suche, W. Sohler, and C. Silberhorn,
 	\RT{Direct generation of genuine single-longitudinal-mode narrowband photon pairs}
 	New J. Phys. {\bf 17}, 073039 (2015).
\bibitem{SPDC.2016a}
	D. Riel{\"{a}}nder, A. Lenhard, M. Mazzera, and H. de Riedmatten,
	\RT{Cavity enhanced telecom heralded single photons for spin-wave solid state quantum memories}
	New J. Phys. {\bf 18}, 123013 (2016).
\bibitem{SPDC.2016b}
	M. Rambach, A. Nikolova, T. J. Weinhold, and A. G. White,
 	\RT{Sub-megahertz linewidth single photon source}
 	APL Photonics {\bf 1}, 096101 (2016).
\bibitem{SPDC.2017a}
	C.-H. Wu, T.-Y. Wu, Y.-C. Yeh, P.-H. Liu, C.-H. Chang, C.-K. Liu, T. Cheng, and C.-S. Chuu,
	\RT{Bright single photons for light-matter interaction}
	Phys. Rev. A {\bf 96}, 023811 (2017).
\bibitem{SPDC.2017b}
	P.-J. Tsai and Y.-C. Chen,
	\RT{Ultrabright, narrow-band photon-pair source for atomic quantum memories}
	Quantum Sci. Technol. {\bf 3}, 034005 (2018).
\bibitem{SPDC.2018}
	K. Niizeki, K. Ikeda, M. Zheng, X. Xie, K. Okamura, N. Takei, N. Namekata, S. Inoue, H. Kosaka, and T. Horikiri,
	\RT{Ultrabright narrow-band telecom two-photon source for long-distance quantum communication}
	Appl. Phys. Express {\bf 11}, 042801 (2018).
\bibitem{SPDC.2019}
	A. Moqanaki, F. Massa, and P. Walther,
	\RT{Novel single-mode narrow-band photon source of high brightness tuned to cesium D2 line}
	APL Photonics {\bf 4}, 090804 (2019).
\bibitem{SPDC.2020}
	J. Liu, J. Liu, P. Yu, and G. Zhang,
	\RT{Sub-megahertz narrow-band photon pairs at 606 nm for solid-state quantum memories}
	APL Photonics {\bf 5}, 066105 (2020).
\end{thebibliography}
\end{document}